# Erich Regener and the maximum in ionisation of the atmosphere

P.Carlson[a] and A.A.Watson[b]

[a]*KTH and the Oscar Klein Centre, Stockholm, Sweden, carlson@particle.kth.se*
[b]*School of Physics and Astronomy, University of Leeds, Leeds, UK, a.a.watson@leeds.ac.uk*

**Abstract**
In the 1930s the German physicist Erich Regener (1881-1955) did important work on the measurement of the rate of production of ionisation deep under-water and in the atmosphere. He discovered, along with one of his students, Georg Pfotzer, the altitude at which the production of ionisation in the atmosphere reaches a maximum, often, but misleadingly, called the Pfotzer maximum. Regener was one of the first to estimate the energy density of cosmic rays, an estimate that was used by Baade and Zwicky to bolster their postulate that supernovae might be their source. Yet Regener's name is less recognised by present-day cosmic ray physicists than it should be largely because in 1937 he was forced to take early retirement by the National Socialists as his wife had Jewish ancestors. In this paper we briefly review his work on cosmic rays and recommend an alternative naming of the ionisation maximum. The influence that Regener had on the field through his son, his son-in-law, his grandsons and his students and through his links with Rutherford's group in Cambridge is discussed in an appendix. Regener was nominated for the Nobel Prize in Physics by Schrödinger in 1938. He died in 1955 at the age of 73.

## 1. Introduction

Erich Regener was born in West Prussia in 1881. He studied physics at the University of Berlin and his 1905 thesis (Regener, 1905) concerns the interaction of UV light on oxygen and ozone molecules and the equilibrium between them, a topic that he revisited on several occasions. In Berlin he chose as the subject for his *Habilitationsschrift* the determination of the elementary electric charge using different methods including alpha-particle scintillations (Regener, 1911). His accurate determination of *e* is very close to today's accepted value.

After a period as professor at the Agricultural University of Berlin and as X-ray field technician during the 1914-18 War, Regener became a full professor of physics at the Technical University of Stuttgart in 1920. A hot topic at this time was the study of atmospheric ionisation following the 1912 discovery of cosmic rays[a] by Victor Hess (1883 – 1964) and from the second half of the 1920s Erich Regener made significant contributions to the study of this field, both through work that he carried out himself and through stimulating work carried out by his son and by his students, both with him and independently: some of these activities are related in appendix A. Regener realised that ionisation measurements had to be made deep in lakes and at the highest altitudes and developed the necessary instruments, along with innovative recording techniques, to make such measurements.

Because his first wife, Viktoria, was a Russian-born Jew Regener had to take "provisional retirement" in 1937 for refusing to divorce her (B. Hoerlin, 2011). While his son and daughter emigrated, Regener started a private research institute on Lake Constance in 1937 that was supported by the Kaiser Wilhelm Society and soon became part of the Society. He was recruited by Werner von Braun to design instruments for rocket flights that were initially to be used to measure the temperature, pressure and density in the upper atmosphere. After the World War II Regener became the first vice-president of the Max Planck Society (formerly the Kaiser Wilhelm Society) and was reinstated to the chair in Stuttgart which he held until his retirement in 1951. In 1952 his private institute was incorporated into the Max-Planck-Gesellschaft as the Max-Planck Institute für Sonnensystemforschung. Regener died in 1955: an obituary written by the Nobel Laureate P M S Blackett appeared in Nature (Blackett, 1959).

---

[a] Before ca 1930 "Höhenstralung" and "Ultrastrahlung" were terms often used for what are now known as "cosmic rays".

**2. Cosmic Ray Work at Stuttgart**

Regener began his cosmic-ray work in Stuttgart in 1928. In that year he noted (Regener, 1928) that the study of the "Hess radiation" is difficult for several reasons: the instrument can contain radioactive material, the air can contain radioactive emanations and the ground contains radioactive radium. Any instrumental radioactivity can be investigated by measuring the ionisation in deep lakes while measurements at high altitudes avoid the ground radioactivity.

Regener was an ingenious experimentalist and soon established a depth-record for observations of the rate of production of ion pairs. He tackled the problem of measuring the decrease in ionisation as a function of increasing depth by submerging a quartz-fibre electrometer and a pressure chamber in the waters of Lake Constance. He succeeded (Regener, 1933a) in measuring the rate of ionisation (pairs of ions produced $cm^{-3} s^{-1}$) down to a depth of 231 m where the rate was found to be very low, of order 0.05 ion pairs $cm^{-3}s^{-1}$. Millikan and Cameron (1928a), working in the US, reached a depth of 50 m where the ionisation was about 2.6 ion pairs $cm^{-3}s^{-1}$. The 'steel bomb' in which Regener measured the ionisation weighed 130 kg and contained $CO_2$ at a pressure of 30 atmospheres. His early work, carried out using a hand-operated winch on a rowing boat, was so successful that it attracted sufficient funding to purchase a motor boat for later work. He named this boat 'Undala', reflecting his belief, and one that was widely held at that time, that cosmic rays were photons with wave-like properties rather than particles. Interesting accounts of the underwater research has been given by Paetzold et al. (1974) and by Pfotzer (1985). Pfotzer worked with Regener, first as a student from 1929, and later, after the war, as a senior scientist.

Regener's work in Lake Constance (Regener, 1931) was highly regarded by eminent physicists of the day including Bruno Rossi (1905 – 1993), a close friend despite their differing views on the nature of cosmic radiation (Rossi, 1985). At the Royal Society discussion meeting of 1931 (Geiger et al., 1931), attended by such as Geiger, Rutherford, C T R Wilson and W H Bragg, both Geiger and Rutherford praised Regener's work although they disagreed as to whether it provided evidence for the particle (Geiger) or wave (Rutherford) nature of the radiation. In Rossi's 1985 article there is a photograph of Rossi with Lise Meitner – also a supporter of the wave hypothesis - and Erich Regener (two of Rossi's "dearest and most respected friends") aboard Regener's boat on the Bodensee. Part of the caption reads "Regener had baptized the boat 'Undula' to reaffirm his faith in the wave nature of cosmic rays. Here, following a discussion on this subject, he was telling me: "If it turns out that you are right, I would have to rename my boat 'Korpuskel', which does not sound as nice as 'Undula'". No date is given for this photograph but it was presumably taken in the early 1930s and before the discovery of the East-West effect that did much to establish the particulate nature of the radiation (Johnson 1933, Alvarez and Compton 1933). A photograph of Regener working aboard the Undula is shown in figure 1.

In the early 1930s Regener started a research program aimed at extending the atmospheric measurements of Hess and of Werner Kolhörster (1881 – 1945) who in 1914 made a flight to 9300 m. Millikan and Bowen (1926) had found it difficult to keep their instruments working at the cold temperatures encountered at high altitudes and Regener overcame this problem by enclosing his equipment in a light gondola encased in cellophane. The cellophane turned the gondola into an effective greenhouse against the low temperatures and the apparatus was operated at what was effectively room temperature.  Again Georg Pfotzer (1909 – 1981) has given a good description of the challenges Regener faced and the solutions that he adopted (Pfotzer, 1985).

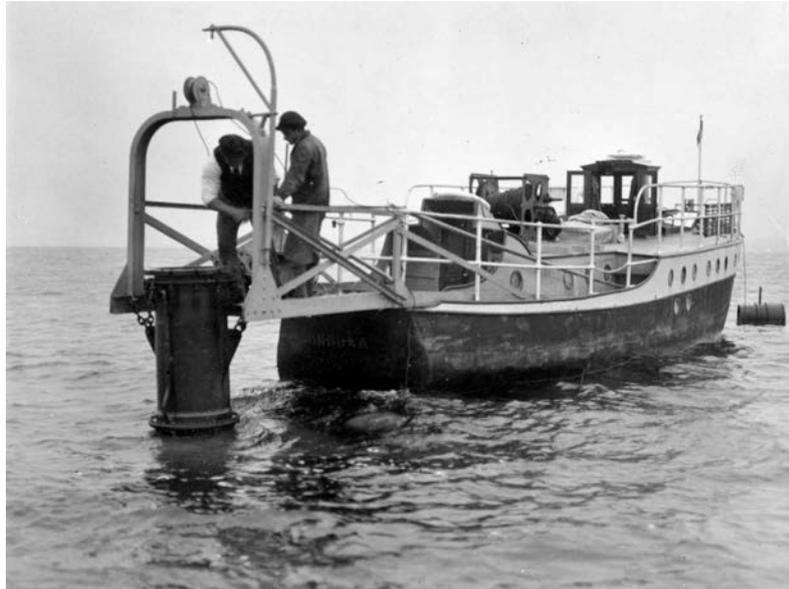

**Figure 1:** Erich Regener, on the left, on board the Undula on Bodensee ~ 1933.
Credit: Archiv der Max-Planck-Gesellschaft, Berlin-Dahlem

**2.1 Results with a balloon-borne electrometer and the energy density of cosmic rays**
Between in 1932 and 1934 Regener made several flights in which he succeeded in making measurements of the rate of ionisation up to heights in the atmosphere where the overburden was less than 50 mm Hg (~ 50 g cm$^{-2}$).  The data from the first flight are shown in figure 2.  The maximum height reached was determined by the point at which one of the three or four lifting balloons burst, ozone interacting with the rubber probably being the cause of the rupture.

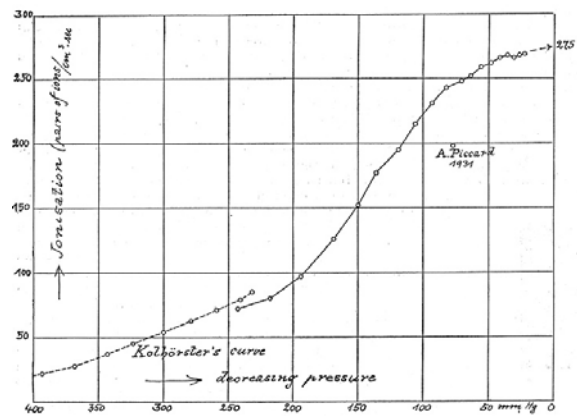

**Figure 2.** Ionisation data from Regener (1932) taken using an electrometer and a 2.1 litre ionisation chamber. The ionisation rate (ion pairs cm$^{-3}$ s$^{-1}$) is shown as function of decreasing pressure (mm Hg). Also shown are data from Kolhörster and Piccard. The extrapolated value is shown as 275 ion pairs cm$^{-3}$s$^{-1}$. As was customary at the time, no errors are shown on the data points. Using these data Regener was able to confirm the results of Kolhörster and infer that the rate of production of ionisation had reached a maximum value which he extrapolated to 275 ion pairs cm$^{-3}$ s$^{-1}$.  He defined this as the 'intensity of the cosmic radiation' at 'its entrance to the atmosphere'.

In a paper submitted on 31 December 1932 (Regener, 1933b) and published only 4 weeks later, he reported the results of integrating the ionisation as a function of height to obtain the total number of ions, produced by the absorption of cosmic rays by a column of air of 1 cm$^2$ cross-section as $1.02 \times 10^8$ ion pairs, to be compared with the value of $1.28 \times 10^7$ ion pairs estimated by Millikan and Cameron (1928b) from a larger extrapolation. Taking 32 eV (about 6% lower than modern measurements) as the energy needed to produce an electron-ion pair in air, he deduced that the energy reaching the earth in the form of cosmic rays was $5.2 \times 10^{-2}$ erg cm$^2$ s$^{-1}$. In a later paper (Millikan et al., 1933) a value of $3.2 \times 10^{-2}$ erg cm$^2$ s$^{-1}$ was derived and compared with the energy falling on the earth as starlight, $6.91 \times 10^{-3}$ erg cm$^2$ s$^{-1}$. Millikan and his co-authors made no reference to Regener's paper although it had been published, in English, nearly 6 months prior to their own work being submitted.

Of course both values for the energy flux are low as both Millikan and Regener were unaware that the bulk of the cosmic rays are charged particles. Both subscribed to the photon view and the values are about a factor two lower than modern estimates. These numbers were used by Baade and Zwicky in their classic paper as evidence to support their theory that cosmic rays were produced in supernovae (Baade and Zwicky, 1934).

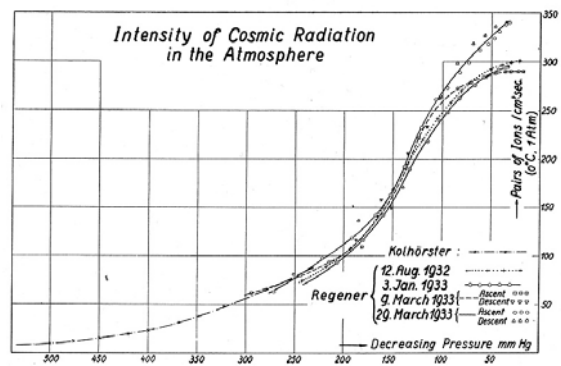

**Figure 3.** Ionisation data from Regener (1933c). The ionisation rate (ion pairs cm$^{-3}$s$^{-1}$) is shown as function of decreasing pressure (mm Hg). Results from 4 flights during 1932-33 with data corrected to 1 atmosphere pressure. Also shown are Kolhörster's data.

From a series of balloon flights in 1932 and 1933 Regener reported the data shown in figure 3. The three lower curves of figure 3 are in good agreement and there is evidence that the rate of production of ion pairs has reached a maximum value. It was normal practice to correct the rate of production of ion pairs to a pressure of one atmosphere: the pressures in the ionisation chamber monitored by the electrometer varied considerably over the range of 3 to 4 atmospheres in different flights.

The results from the flight of 29 March 1933 are of particular interest. The instrument was apparently working well and no instrumental reason could be identified to account for the much higher rate of ionisation. Regener considered the possibility that radioactivity from the moon might be responsible. He rejected this as he did the possibility that 'a magnetic disturbance of medium strength' that occurred on that day, while it was magnetically calm on the other days, had an influence. He noted that "it would be remarkable if there were a connexion between the magnetic intensity and the intensity of cosmic rays in the highest part of the atmosphere, and only in the highest parts: that is to say, that there are additional rays (soft rays) there perhaps coming from a sunspot". Of course with the benefit of hindsight one might remark that the solar connection should have perhaps have been given more attention but rather the statement demonstrates the strong adherence that Regener, and others, had to the photon hypothesis. Indeed he discusses the shape of the curve at the highest altitudes as being evidence for an electromagnetic component.

**2.2 Discovery of the maximum**

In his 1933 Nature paper (Regener, 1933c) Regener noted that the ionisation data is for rays coming from all directions. He then went on to report that his collaborator, Bernhard Gross, had calculated the ionisation data for vertically incident rays showing that the intensity diminished towards the top of the atmosphere i.e. *shows a maximum*. In that paper (Gross, 1933), received on 6 April 1933 and published 14 June 1933, Gross acknowledges Regener for suggesting the problem and for his help with the work. The results of Gross's calculations are shown in figure 4.

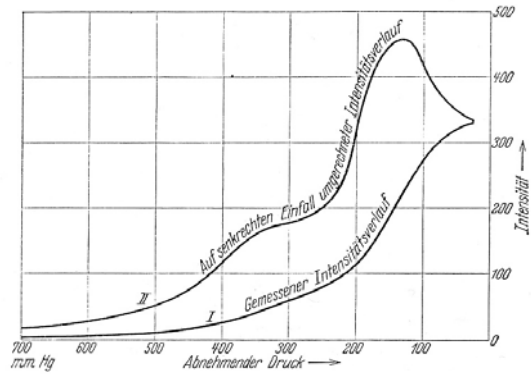

**Figure 4.** The ionisation rate (ion pairs $cm^{-3}s^{-1}$) as function of decreasing pressure (mm Hg) from the work of Gross (1933). Line I shows the measured rate (Regener data) for rays from all directions. Line II is the calculated rate for vertically incident rays showing a clear maximum at about 130 mm Hg and a shoulder at 350 mm Hg.

As an excellent experimentalist, Regener wanted to exploit the Geiger-Muller-tubes, developed in 1928, so that he could register single charged particles. He involved Georg Pfotzer, one of his students, in this work. Their first results, obtained with one tube carried to an altitude of 28 km, were published in September 1934 (Regener and Pfotzer, 1934) with the ionisation measured found to be the same above 18 km height as reported earlier with an ionisation chamber.

A year later, in November 1935, Regener and Pfotzer described the results from two ascents with three GM tubes in coincidence, arranged to measure the vertical intensity. The solid angle was 20 degrees about the zenith (Regener and Pfotzer, 1935). Data show a clear maximum for 100 mm Hg pressure (~16 km) and a bump at 300 mm Hg, see figure 5. Note the author-ordering in these two Nature papers.

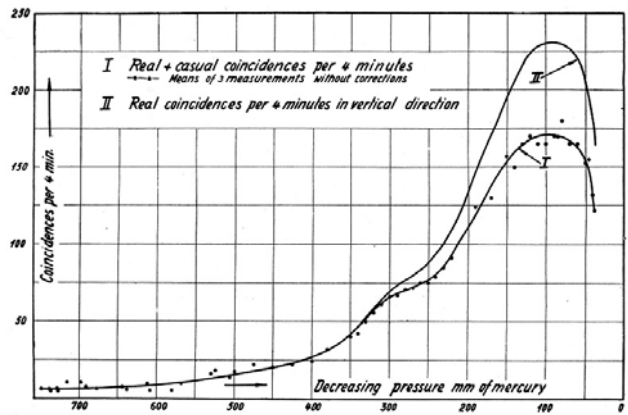

**Figure 5.** Data from coincidence measurements by Regener and Pfotzer (1935). The coincidence rate per 4 minute in a solid angle of 20 degrees about the zenith is shown as function of decreasing pressure (mm Hg) (data points and line I). Data, corrected for dead-time losses, are shown as line II. Notice the clear maximum at about 100 mm Hg.

Pfotzer gives a detailed account of this work together with results from other flights in his 1936 thesis, presented on 15 May, 1936 and in part published in Zeitschrift f. Physik (Pfotzer, 1936). In the acknowledgements Pfotzer thanks Regener warmly: *"My sincere thanks to my idolized teacher prof. Regener for organizing the ascents and for continuous support and help."*

Pfotzer went on to have a distinguished career in physics, following Regener as Director of the Max Planck Institute für Stratosphärenforschung in 1956. However, it is clear to us that the real discoverer of the maximum in the rate of production of ion pairs with altitude, misleadingly called the Pfotzer maximum, was Erich Regener with the 1935 result essentially confirming those shown in figure 4.

### 3. Why not the Regener maximum or even the Regener-Pfotzer maximum?

It is not at all clear why Regener's name is not now associated with the maximum in the altitude dependence of the rate of production of ion pairs. In their classic paper (Bhahba and Heitler, 1937), submitted in December 1936, Bhabha (1909 – 1966) and Heitler (1904 – 1981) use the work of Regener and Pfotzer, along with the Rossi transition curves (Rossi, 1938), as demonstrations of the accuracy of their theory of cosmic showers. Phrases such as 'Regener's measurements of the absorption of cosmic rays in the atmosphere' are used and in the discussion of comparisons of theory with experimental data (section 7B) the title of the section is 'The Regener Curve at High Altitude'. Although Pfotzer's 1936 paper is in the reference list, and, curiously, the order of the authors in the citation of Regener and Pfotzer (1935) has Pfotzer's name first, it is clear that Regener is seen by no less authorities than Bhabha and Heitler as the discoverer of the maximum. Heitler must surely have known what was going on in Germany and certainly Bhabha, at that time working in Cambridge in Rutherford's laboratory, will have been aware of the close links of Rutherford and Blackett to Regener (Carmichael, 1985). In a paper written in 1938, Rossi (Rossi, 1938) speaks of the Regener-Pfotzer curve at 100 g cm$^{-2}$: no citation is given. However in 1941 Schein, Jess and Wollan (Schein et al., 1941) showed a plot of the ionisation as a function of altitude in which the maximum is labelled *'the Pfotzer maximum'*. In this paper, and many subsequent papers, the maximum is generally and, we contend, misleadingly referred to as the Pfotzer maximum, without citation.

It is our speculation that Regener advised Pfotzer against joint-authorship as by 1936 his difficulties with the National Socialist party were increasing. A fascinating account of these difficulties, based on letters between her parents, has been given by the daughter of one of Regener's students (B. Hoerlin, 2011). In 1933, during the period of book-burning in Germany, Regener, along with the Jewish librarian in Stuttgart, were rumoured to be targets of angry students supportive of the National Socialists. Accordingly Hermann Hoerlin appointed himself as guardian of Regener and his family and, armed with a wooden ice axe, slept on the floor behind the door of Regener's apartment. Bettina Hoerlin also notes that Regener was one of 75 established physicists (including Heisenberg) who in October 1936 sent a petition to the Reich's Minister of Education warning of the declining state of German physics. She remarks that 'the contrarian act of signing the petition' – along with the issue of his wife – 'sealed Regener's removal from the Stuttgart Faculty'.

Regener was nominated by Schrödinger for the Nobel Prize in Physics in 1938 for his detailed measurements of ionisation rates, both in the atmosphere and deep under water.

It is our firm conclusion that the atmospheric and cosmic-ray communities should start talking and writing of the Regener-Pfotzer maximum, or even simply the Regener maximum, when discussing the position in the atmosphere at which the rate of production of ionisation becomes a maximum.

**References**

Alvarez, L. and Compton, A.H., 1934. A Positively Charged Component of the Cosmic Rays. Physical Review 43, 835-836, 1933

Baade, W., Zwicky, F., 1934. Cosmic Rays from Super-Novae. Proc. Nat. Acad. USA 20, 259-263.

Bethe, H.A., Kerr, D.M., Jeffries, R.A., 1984. Obituary H.W. Hoerlin. Physics Today Dec. 1984, pp.82-83.

Bhabha, H.J. and Heitler, W., 1937. The Passage of fast Electrons and the Theory of Cosmic Showers. Proc Roy Soc A 159, 432-458.

Blackett, P.M.S., 1959. Obituary: Erich Regener. Nature 175, 1107-1108.

Carmichael, H., 1985. Edinburgh, Cambridge, and Baffin Bay, in: Sekido, Y., Elliot, H. (Eds), Early History of Cosmic Ray studies. D. Reidel 1985, pp. 99-113.

Carmichael, H., Dymond, E.G., 1938. High-Altitude Cosmic Radiation Measurements near the Magnetic Axis-Pole. Nature 141, 910-911.

Clay, J., Berlage, H.P., 1932. Variation der Ultrastrahlung mit der geographischen Breite und den Erdmagnetismus. Naturwiss. 20, 687-688. (In German) (Variation of the ultra-radiation with latitude and with the earth's magnetic field).

Compton, A.H., 1933. A Geographic Study of Cosmic Rays. Phys. Rev. 43, 387-403, 1933

Geiger, H., Lord Rutherford, Regener, E. et al., 1931. Discussion on Ultra-Penetrating Rays. Proc. Roy. Soc. A 132, 331-352.

Greisen, K., 1966. Highlights in Air Shower Studies, 1965. Proceedings of the 9[th] International Conference on Cosmic Rays, Institute of Physics, Volume 2, pp. 609-615

Gross, B., 1933. Zur Absorption der Ultrastrahlung. Zeit. Phys. 83, 214-221. (In German: On the absorption of the Ultra-Radiation).

Hess, V.F., 1912. Über Beobachtungen der durchdringenden Strahlung bei sieben Freiballonfahrthen. Physicalische Zeitsschfirft 13 1084-1091 (In German: On the observation of the penetrating radiation from seven balloon flights).

Hoerlin, Bettina, 2011. 'Steps of Courage' *My Parents' Journey from Nazi Germany to America,* p 49 and p 130, Author House, USA.

Hoerlin, H., 1933. Latitude Effect of Cosmic Radiation. Nature 132, 61-61.

Hoerlin, H., 1963 Air Fluorescence Excited by High Altitude Nuclear Explosions. Los Alamos Report, LA-3417-MS 1- 74.

Johnson, T.H., The Azimuthal Asymmetry of the Cosmic Radiation. Physical Review 43, 834-835, 1933

Kolhörster, W. 1914. Messungen der durchdringenden Stralungen bis in Höhen von 9300 m, Verh. deutsche phys. Gesellschaft 16, 719–721. (In German: Measurements of the penetrating radiation up to 9300 m)

McCusker, C.B.A. Winn, M.M., Rathgeber, H.D., 1963. Two Large Air Shower Experiments, Proceedings of 8th International Conference on Cosmic Rays, Jaipur, Volume 4 pp. 306–309.

Mead, J.B., 1963. Properties of Teller Light (Air Fluorescence) induced by 22 MeV electrons, CRD Sigma 3, UCRL-7604, Lawrence Radiation Laboratory (paper remains classified, 2014).

Millikan R.A., Bowen I.S., 1926. High Frequency Rays of Cosmic Origin I. Sounding Balloon Observations of Extreme Altitudes. Phys. Rev. 27, 353-361.

Millikan, R.A., Cameron, G.H., 1928a. New Results on Cosmic Rays. Nature 121, 19-26.

Millikan R.A, Cameron, G.H., 1928b. New Precision in Cosmic Ray Measurements; Yielding Extension of Spectrum and Indications of Bands. Phys. Rev. 31, 921-930.

Millikan R.A., Bowen I.S., Neher H.V., 1933. New High-Altitude Study of Cosmic-Ray Bands and a new Determination of Their Energy Content. Phys. Rev. 44, 246-252.

Paetzold, H.K., Pfotzer, G., Schopper, E., 1974. Erich Regener als Wegbereiter der extraterrestrischen Physik, in: Birett, H., Helbig, K., Kertz, W., Schmucker, U. (Eds), Zur Geschichte der Geophysik . Springer Berlin pp. 167-188. (In German) (Erich Regener a pioneer in extraterrestrial radiation physics, in: Geophysics history).

Pfotzer, G., 1936. Dreifachkoinzidenzen der Ultrastrahlung aus vertikaler Richtung in der Stratosphäre. Zeit. Phys. 102, 23-58. (In German) (Vertical Intensity of Cosmic Rays in the Stratosphere by Threefold Coincidences).

Pfotzer, G., 1985. On Erich Regener's cosmic-ray work in Stuttgart and related subjects, in: Sekido, Y., Elliot, H. (Eds), Early History of Cosmic Ray Studies. D. Reidel 1985, pp. 75-89.

Regener, E., 1905. Über die chemische Wirkung kurzwelliger Strahlung auf gasförmige Körper, thesis univ. Berlin (In German). (Short wave-length interactions in gases). Also Ann. d. Physik 20, 1033-1046, 1906.

Regener, E., 1911. Über Ladungsbestimmungen an Nebelteilchen. Phys. Zeit. 12, 135-141. (In German). (Estimating the charges on droplets).

Regener, E., 1928. Über neuere Versuche über die sogennante durchdringende Höhenstrahlung in der Erdatmosphäre. Naturwissentschaftlige Monatshefte 25, 240. (In German). (New experiments on the penetrating rays in the atmosphere).

Regener, E., 1931. Spectrum of Cosmic Rays. Nature 127, 233-234.

Regener, E., 1932. Intensity of Cosmic Radiation in the High Atmosphere. Nature 130, 364-364.

Regener, E., 1933a. Die Absorptionskurve der Ultrastrahlung und ihre Deutung. Phys. Z. 34, 306-323 (In German) (Absorption curve of the penetrating radiation and its interpretation).

Regener, E., 1933b. Energy of Cosmic Rays. Nature 131, 130-130.

Regener, E., 1933c. New Results in Cosmic Ray Measurements. Nature 132, 696-698.

Regener, E., Regener, V.H., 1934. Aufnahme der ultravioletten Sonnenspektrums in der Stratosphäre und vertikale Ozonverteilung. Phys. Zeit. 35, 788-793. (In German) (Measurement of the ultraviolet spectrum from the sun and the vertical ozone distribution).

Regener, E., Pfotzer, G., 1934. Messungen der Ultrastrahlung in der oberen Atmosphäre mit dem Zählrohr. Phys. Zeit. 35, 779-784. (In German). Intensity of the Cosmic Ultra-Radiation in the Stratosphere with the Tube-Counter. Nature 134, 325-325, 1935.

Regener, E., Pfotzer, G., 1935. Vertical Intensity of Cosmic Rays by Threefold Coincidences in the Stratosphere. Nature 136, 718-719.

Regener, V.H., 1951. Statistical Significance of Small Samples of Cosmic-Ray Counts. Phys. Rev. 84, 161-162.

Regener, V.H., Regener, E., 1970. Cosmic Rays Underground and the Interplanetary Magnetic Field., Zeitschrift für Geophysik 40, 761 – 766.

Rossi, B., 1933. Über die Eigenschaften der durchdringenden Korpuskularstrahlung im Meerniveau. Zeit. Phys. 82, 151-178. (In German) (Properties of the penetrating particle radiation at sea level).

Rossi, B., 1938. Further Evidence for Radioactive Decay of Mesotrons. Nature 142, 993-993.

Rossi, B., 1985. Arcetri, 1928 – 1932, in: Sekido, Y., Elliot, H. (Eds), Early History of Cosmic Ray Studies. D. Reidel 1985, 53-89

Schein, M., Jesse, W.P, Wollan, E.O., 1941. The Nature of the Primary Cosmic Radiation and the Origin of the Mesotron. Phys. Rev. 59, 615-615.

Swinson, D.B. , 1995. Diurnal variations underground since 1959. Proceedings of the 24$^{th}$ International Cosmic Ray Conference, Rome, Vol 4, pp. 627-630.

**Appendix A: Regener's influence through his relations, students and other contacts**

In addition to the work which Regener led himself, he had significant – but little recognised – influence on cosmic ray research in a number of areas. His student Hermann Hoerlin, a keen climber, with support from the *Deutscher und Oesterricher Alpverein,* made measurements of cosmic ray ionisation in the Peruvian Andes at altitudes up to 6100 m. On the outward and return voyages he was able to make extensive measurements of the cosmic ray intensity as a function of latitude. Data taken on the homeward voyage from the Straits of Magellan to Hamburg form a remarkable record (Hoerlin, 1933) and confirmed the results of Clay and Berlage (1932) and of Compton (1933) obtained only slightly earlier: Hoerlin's work preceded that of Millikan. Hoerlin too married a woman with Jewish ancestry and they emigrated to the United States in 1938. There Hoerlin continued working for AGFA, the German manufacturers of photographic material and equipment with whom he had found a job in Germany after completing his PhD. In 1953, with the encouragement of Hans Bethe, he moved to Los Alamos where he headed the Nuclear Weapons Effects group. After the war he studied the fluorescence emission from nuclear explosions (Hoerlin, 1963) and was regarded as one of the world's experts on high-altitude nuclear detonations (Bethe et al., 1984). This method of explosion monitoring was an idea of Teller's (Mead, 1963) and may have been an inspiration for Greisen's proposal of the fluorescence technique for the observation of the highest energy cosmic rays. Hoerlin was deeply devoted to Regener, as an example cited in section 3 of the main text illustrates.

Regener's son, Victor (1913 – 2006) emigrated to the United States before the start of World War II to work with Compton in Chicago. He moved to the University of New Mexico in Albuquerque and established a well-known cosmic ray group there (Swinson, 1995) dedicated to the study of the time variations of low energy cosmic rays. He had made an early foray into cosmic ray work when he accompanied Hoerlin on an expedition to the Jungfraujoch in 1933 while, with his father, he studied ozone at high altitudes (Regener and Regener, 1934). Victor Regener had a strong interest in statistics and was one of the first physicists to address the question of how to assign uncertainties to small samples in a paper entitled *Statistical Significance of Small Samples of Cosmic Ray Count* (Regener, 1951). Later with his son, Eric, a mathematician and a musician, Victor Regener developed a mathematical analysis of importance in establishing correlations between interplanetary magnetic field parameters and cosmic ray diurnal variations as measured with underground data (Regener and Regener, 1970).

In 1932 Regener's daughter, Erika, married one of his research students, Henri Rathgeber. Rathgeber completed his undergraduate work and doctoral work at the University of Stuttgart. The Rathgebers emigrated to Australia in 1938 where he was employed first in Melbourne on munitions work during World War II and later at the University of Melbourne where he studied cosmic rays. In 1952 he moved to Sydney working on cosmic rays with H Messel and later with C B A McCusker. With McCusker he developed a technique to study the structure of the cores of extensive air showers using large areas of liquid scintillator and image intensifiers (McCusker, Winn and Rathgeber, 1963). Rathgeber's son, Michael, was very skilled at electronics and played an important role in McCusker's air shower work until his early death in 1969. A scholarship was established at the University of Sydney by his parents in his name in 1971.

Erich Regener was highly regarded by both Rutherford and Blackett. Blackett wrote his obituary (Blackett, 1959) and Rutherford sent one of his students, H Carmichael, to Stuttgart to be trained by Regener in the art of flying balloons. Carmichael has given an interesting account of this visit (Carmichael, 1985) which took place before an observing campaign that he made with Dymond to Baffin Island on the Wordie Expedition in 1937 (Carmichael and Dymond, 1938).

**Contributions of Authors**
The authors have contributed equally to the writing of this paper.

**Acknowledgements**
AAW would like to thank Virginia Trimble for pointing out the use of Regener's data on the energy density of cosmic rays by Baade and Zwicky. We are grateful to Derek Swinson for drawing our attention to his 1995 paper.